\documentstyle[preprint,aps,tighten]{revtex}
\begin{document}
\draft
\title{Critical level spacing distribution of two-dimensional
     disordered systems with spin-orbit coupling}
\author{L. Schweitzer$^1$ and I. Kh. Zharekeshev$^2$\cite{byline}}
\address{$^1$Physikalisch-Technische Bundesanstalt, Bundesallee 100,
D-38116 Braunschweig, Germany\\
$^2$1.~Institut f\"ur Theoretische Physik, Universit\"at Hamburg,
Jungiusstrasse 9, D-20355 Hamburg, Germany}
\maketitle
\begin{abstract}
The energy level statistics of 2D electrons with spin-orbit scattering
are considered near the disorder induced metal-insulator transition.
Using the Ando model, the nearest-level-spacing distribution
is calculated numerically at the critical point.
It is shown that the critical spacing distribution is size independent
and has a Poisson-like decay at large spacings as distinct from the
Gaussian asymptotic form obtained by the random-matrix theory.
\end{abstract}

Recently, the statistical description of electronic properties at the
critical region near the Anderson transition in disordered systems has
been the subject of great interest. Universal behavior was
proposed~\cite{Altshuler88,Shklovskii93} for the variance of
the ensemble averaged number of energy levels in a given interval,
$\langle (\delta N)^{2}\rangle = \langle N^{2}\rangle -\langle N\rangle^{2}$,
and for the nearest-level-spacing distribution $P(s)$.
A relation $\langle (\delta N)^{2}\rangle =a\,\langle N\rangle$ could be
extracted from numerical calculations for the
three-dimensional Anderson model,
while for $P(s)$ a universal combination of the well known Wigner surmise
(small spacings) and the Poisson distribution (large spacings) was proposed.
Recent analytical calculations~\cite{Aronov94,Kravtsov94} suggested, however,
another universal relation $\langle (\delta N)^{2}\rangle =a_{d\beta }
\langle N\rangle^{\gamma}/\beta$ near the critical transition (mobility edge)
and the corresponding $P(s)$ was different from both the Poisson and the
Wigner limit.
Universal means here that the result depends only on the symmetry class
$\beta $, the correlation exponent $\nu = [d(1-\gamma)]^{-1}$, and the
spatial dimension $d$.
The statistics of energy levels are governed by the symmetry of the
Hamiltonian belonging to certain universality classes of the corresponding
Gaussian ensembles of random matrices:
orthogonal ($\beta =1$), unitary ($\beta =2$) and
symplectic ($\beta =4$)~\cite{RMT}.

In two-dimensional (2D) disordered systems a complete Anderson
transition is expected only for the symplectic symmetry class.
Two different models for describing the
localization problem in two dimensions
in the presence of spin-orbit coupling were proposed~\cite{Evangelou87,Ando88}.
For the Ando model~\cite{Ando89} the critical
behavior of the localization length
was previously studied by means of the transfer-matrix method
\cite{Fastenrath91,Fastenrath92,Markos94}.
The statistics of energy levels for symplectic symmetry in connection with
the results of the random-matrix theory (RMT) was considered in the
Evangelou-Ziman model~\cite{Evangelou89} and the distribution $P(s)$
at the transition point
for this model has already been mentioned~\cite{Evangelou90}.
The multifractal properties of the critical eigenstates in the
symplectic case were investigated in~\cite{Chalker93,Schweitzer95}.

In this paper we study the level statistics of electrons
with spin-orbit coupling in 2D disordered systems.
By using the Ando model we calculate numerically the correlations
in the exact one-electron energy spectrum
at the metal-insulator transition.
Our main result is that the probability density $P(s)$ of neighboring
levels in the critical region is universal, i.e. it does not depend on
the size of the system, and has a novel form.

It is known that in the metallic phase $P(s)$ is very close
to the Wigner surmise~\cite{Wigner55} appropriate for the symplectic
symmetry class of the random Hamiltonians~\cite{Efetov83},
\begin{equation}
P_{\rm GSE}(s)=\frac{2^{18}}{3^6\,\pi^3}\,s^4\,\exp\left(-\frac{64}{9\,\pi}
s^2 \right ),
\label{GSE}
\end{equation}
where $s$ is measured in units of the mean level spacing.
In the strongly localized regime the energy spectrum is
completely uncorrelated, and
the distribution of the level spacings obeys the Poisson law
\begin{equation}
P_{\rm P}(s) = \exp(-s).
\label{Poisson}
\end{equation}

In addition to these two universal distributions one can expect that
there is a third form of $P(s)$, which
corresponds exactly to the metal-insulator transition.
This new universal statistics was already found for the
3D disordered systems without spin-orbit scattering ($\beta=1$) and
confirmed numerically~\cite{Shklovskii93,Hofstetter94}.
An analogous result was recently obtained in the unitary case
($\beta=2$)~\cite{HofstetterS94}, where time-reversal symmetry is broken
by the magnetic field.

It is reasonable to suppose that a similar scale-invariant universality
of $P(s)$ also holds at the critical point for 2D disordered electrons
with spin-orbit coupling.
Thus, we expect that in the thermodynamic limit
there exist three possible limiting situations for $\beta=4$
namely, the Wigner surmise Eq.~(\ref{GSE}) for the metallic regime,
the Poisson law Eq.~(\ref{Poisson}) for the insulating regime,
and the critical distribution $P(s)$ at mobility edge.
Therefore, if $L\to \infty $, the level statistics change discontinuously
twice, from the delocalized regime to the transition point and, then, from
the transition point to the localized regime.

In order to calculate the critical level statistics we start with
the Hamiltonian of the Ando model~\cite{Ando89}
\begin{equation}
H=\sum_{n,\sigma }\epsilon_{n}^{} c_{n \sigma }^{\dag} c_{n \sigma }^{} +
\sum_{n,m,\sigma,\sigma '}
V(n,\sigma ;m,\sigma ')\,c_{n \sigma }^{\dag} c_{m \sigma'}^{},
\label{Hamil}
\end{equation}
where $c_{n \sigma }^{\dag}$ and $c_{n \sigma }^{}$ are the creation
and annihilation operators of
an electron at a lattice site $n=(x,y)$ with the spin component $\sigma $, and
$m$ denotes the sites adjacent to the site $n$.
The on-site energy $\epsilon_{n}$
is randomly distributed around zero according to a box distribution with
a width $W$.
The parameter $W$ specifies the degree of the disorder.
The transfer matrices $V(n,\sigma ;m,\sigma ') = V_{x}, V_{y}$ depend on the
direction
\begin{equation}
V_{x,y;x+1,y}=\left( \begin{array}{cc}
V_{1} & V_{2} \\ -V_{2} & V_{1} \end{array} \right), \qquad
V_{x,y;x,y+1}=\left( \begin{array}{cc}
V_{1} & -iV_{2} \\ -iV_{2} & V_{1} \end{array} \right)
\label{spinor}
\end{equation}
and describe the hopping between the nearest neighbor sites in the lattice.
The strength of the spin-orbit coupling is given by the parameter
$S=V_{2}/V$, where $V=(V_{1}^{2}+V_{2}^{2})^{1/2}$ is taken as the unit of
energy. In what follows we consider $S=1/2$.
It was earlier found by the transfer-matrix method~\cite{Fastenrath92} that
the metal-insulator transition in the middle of the band $\varepsilon/V=0$
occurs at a disorder $W_{c}/V=5.74$.

Applying periodic boundary conditions, the exact discrete eigenvalue
spectrum has been obtained from a numerical diagonalization of the
Hamiltonian Eq.~(\ref{Hamil}) using a Lanczos algorithm.
For square lattices of linear size $L/a=50$ (100) energy intervals
[$-1,0$] ([$-0.5,0$]) were chosen. Within these intervals the density of
states is almost constant, $\rho = (\Delta L^{2})^{-1}\approx 0.127$.
Here, $a$ is the lattice constant and $\Delta $ the mean level spacing.
The eigenvalues taken from these intervals belong to the critical region
where the correlation length $\xi$ is larger than the system size $L$.
The total numbers of eigenvalues $N$ were
94672 (300 realizations) and 101744 (160 realizations) for $L=50\,a$ and
$L=100\,a$, respectively.

Fig.~\ref{fig1}\, displays the level spacing distribution function
$P(s)$ calculated for the
different sizes of the system at the metal-insulator transition.
One can see that independent of the system size
the data lie on a common curve.
This critical $P(s)$ is very close to $P_{\rm GSE}$
and passes the point $s_{0} \approx 1.63$,
where the two limiting distributions $P_{\rm GSE}(s)$ and $P_{P}(s)$ cross.
For small spacings $P(s) \propto s^4$.
It is interesting to notice that our results differ from those obtained
previously for smaller systems of the Evangelou-Ziman spin-orbit coupling
model~\cite{Evangelou90}.

We applied the fitting function which was recently proposed for the
description of the critical level distribution~\cite{Aronov94},
\begin{equation}
P(s) = B\,s^{4}\,\exp(-A\,s^{2-\gamma})
\label{fit}
\end{equation}
and found $A = 2.77 \pm 0.05$, $B = 17.8\pm 0.8$ and $\gamma = 0.28\pm 0.03$.
Using a confidence level of 95\% ($\alpha=0.05$) the fitting procedure within
the range $0<s<3$ where the statistics of our numerical data is rather good
yielded $\chi^2 = 29.2$ which is less than the expected value
$\chi_{\alpha}^2 = 63.4$. Hence, the analytical formula (Eq.~\ref{fit})
can be accepted within an approximate relative error of $(\chi^{2}/N)^{1/2}
\approx 2$\,\%.
Although, in the range $0<s<3$ the calculated P(s) is in good agreement
with Eq.~(\ref{fit}), the obtained exponent $\gamma$ gives, however, a
different value of the correlation length exponent $\nu = [(1-\gamma)\,d]^{-1}
\approx 0.7$ as compared to the numerically obtained value $\nu=2.75$ using
the transfer-matrix method~\cite{Fastenrath91,Fastenrath92}.

Moreover, for large spacings our results deviate
markedly from the above Eq.~(\ref{fit}) (inset of Fig.~\ref{fig1}).
Instead, the behavior of $P(s)$ in the range of the spacings $1.5<s \lesssim 4$
is well described by the Poisson-like asymptotic form
\begin{equation}
P(s)\propto \exp(-\kappa s),
\label{crit}
\end{equation}
with a coefficient $\kappa = 4.0\pm 0.2$.
This decay is much slower than both the Gaussian decay
of Eq.~(\ref{GSE}) and the intermediate decay from
Eq.~(\ref{fit}), but faster than that for the insulating regime
Eq.~(\ref{Poisson}).
A similar exponential tail of the critical $P(s)$ was found
also in the 3D-case
without spin-orbit interactions~\cite{ZharekeshevK95}
for which the coefficient $\kappa $ is approximately a factor of 2 less.
Such an asymptotic behavior of the critical $P(s)$
is in good agreement with a suggestion
about the Gaussian form for the distribution of the number of levels
lying within a given energy interval~\cite{Altshuler88,Shklovskii94}.

In order to diminish the magnitude of relative fluctuations due to
the limited number of realizations and to analyze the asymptotic
behavior of the level spacing distribution in detail,
it is more convenient to consider a total probability function
$I(s)=\int_{s}^{\infty} P(s^{\prime})\,ds^{\prime}$.
This quantity implies a portion of spacings which are larger
than a given $s$.
It is clear that $I(0)=1$ and
$\int_{0}^{\infty} I(s^{\prime})\,ds^{\prime}= 1$ regardless of the
disorder. In the strongly localized regime $I_{\rm P}(s)=\exp(-s)$, and
$I_{\rm GSE}(s)$ can be calculated from the RMT.
The results of the numerical calculations
for the critical $I(s)$ are shown in Fig.~\ref{fig2}.
One observes again a discrepancy from both the GSE asymptotic and
the $I(s)$ obtained from Eq.~(\ref{fit}), particularly when $s$ is large.
The form of the critical $I(s)$ is not sensitive
to the change of the lattice size, in analogy with $P(s)$.
The Poisson tail as described by Eq.~(\ref{crit}) is recovered
for large spacings: $\ln I(s) \propto -s$.
We have checked that the slope of the linear behavior of $\ln I(s)$
at the transition does not depend on the width of the energy
interval from which the levels are taken, as long
as they belong to the critical region.
In the range of very small probability ($s>3$)
the larger fluctuations observed are due to insufficient statistical data.
However, the accuracy of the total calculated $I(s)$ is higher than that
of $P(s)$. One should notice that the behavior of the obtained level spacing
distribution does obviously not reflect the information on the critical
exponent $\nu$ and the dimensionality $d$ in the form as it is expressed
by Eq.~(\ref{fit}).

In conclusion, we have presented results of computer simulations of the
nearest-level-spacing distribution $P(s)$ for 2D disordered systems
in the presence of spin-orbit interaction.
Exactly at the metal-insulator transition the $P(s)$ and, consequently,
the total probability of neighboring spacings $I(s)$ do not depend on the
system size and are different from the universal limiting
distributions corresponding to the metallic and the insulating regime.
They appear to exhibit critical
behavior at the disorder $W_{c}/V=5.74$ and finite-size scaling properties
around the critical point.
The obtained large-$s$ parts of $P(s)$ and $I(s)$ are shown to have a
Poisson-like decay so that $\ln P(s) = -\kappa\,s$ where $\kappa \approx
4.0$ is larger as compared to the insulating regime ($\kappa =1$).

We are grateful to B.~Kramer and B.~Shklovskii for helpful discussions.
I.Kh.Zh. thanks the DAAD for financial support during his stay at the
University of Hamburg.

\bibliographystyle{prsty}

\begin{figure}[htb]
\caption[]{The level spacing distribution $P(s)$ for two different system
sizes at the critical disorder $W_{c}/V=5.74$.
Solid lines correspond to $P_{\rm GSE}(s)$ and $P_{\rm P}(s)$,
respectively. The dotted line is a fit of Eq.~(\ref{fit}).
The inset shows the large-$s$ behavior of $P(s)$.
The straight line is a fit according to Eq.~(\ref{crit}).
}
\label{fig1}
\end{figure}

\begin{figure}[htb]
\caption[]{The integrated probability $I(s)$ at the critical disorder
$W_{c}=5.74$.
The solid curves are $I_{\rm GSE}(s)$ and $I_{\rm P}(s)$
for the metallic and insulating phases, respectively.
The dotted line is obtained from Eq.~(\ref{fit}).
The inset shows the large-$s$  part of $I(s)$.
The straight line fitting the data is $\ln I(s)= -4.0\,s+3.6$.
}
\label{fig2}
\end{figure}
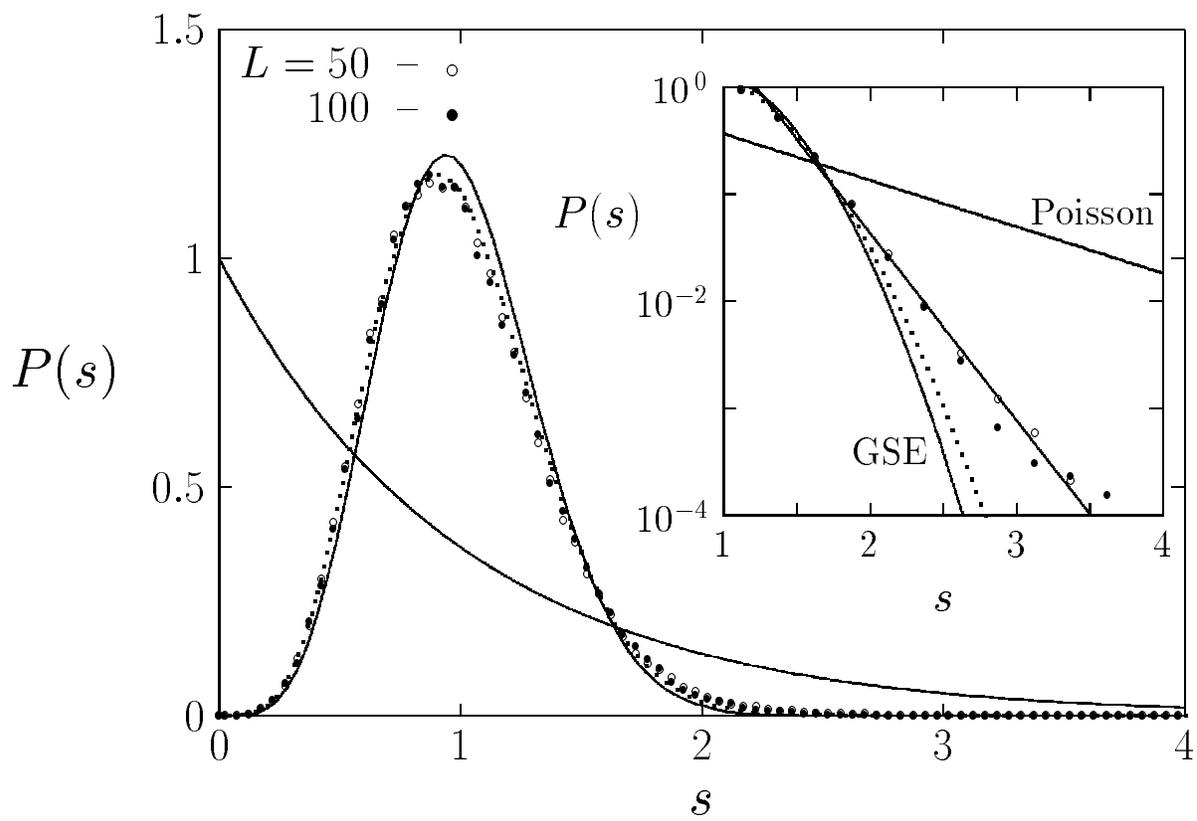
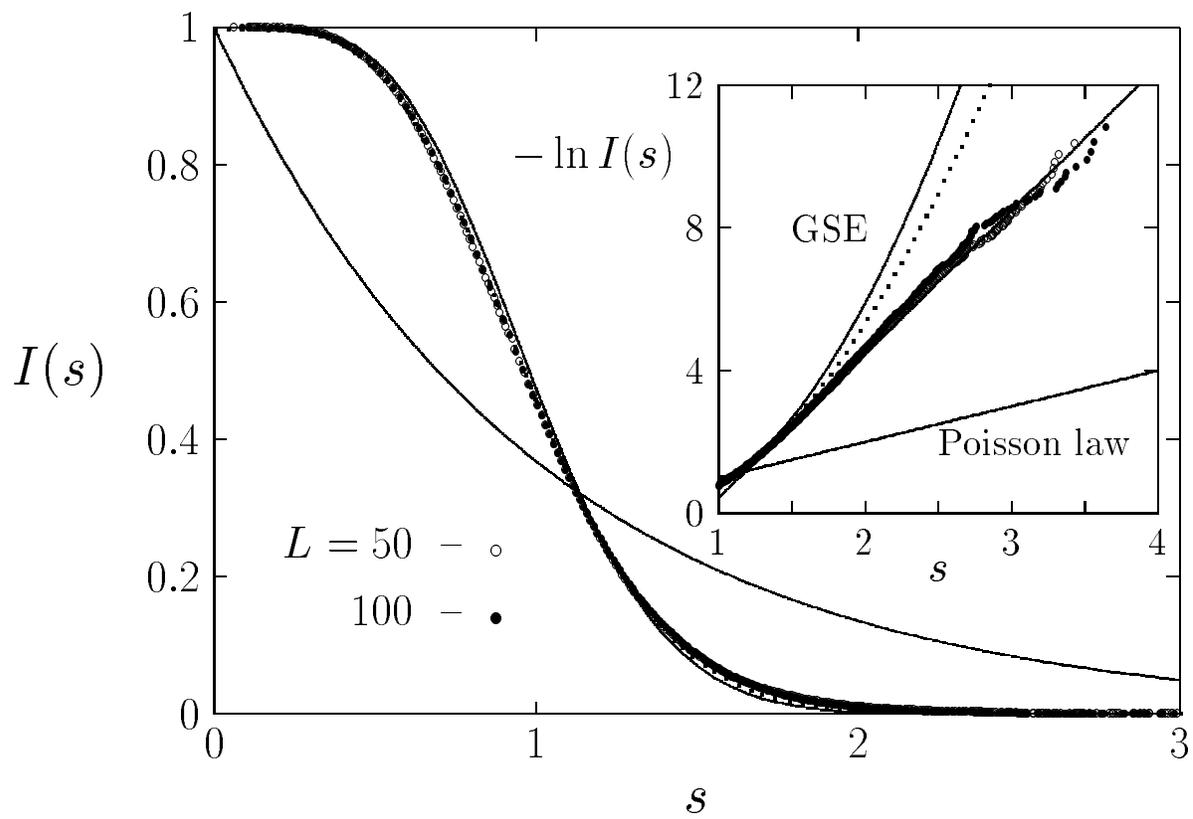
\end{document}